\documentclass[11pt,twocolumn,twoside,a4paper,amsmath,amssymb,aps,showkeys,showpacs]{revtex4}

\usepackage{amsfonts}
\usepackage{graphics} 
\usepackage{epsfig}
\usepackage{fancyheadings}

\newcommand{\p}[1]{(\ref{#1})}

\begin{document}
\thispagestyle{myheadings}
\rhead[]{}
\lhead[]{}
\chead[A.A. Isayev, J. Yang]{Spin polarization phenomena in dense neutron matter}

\title{Spin polarization phenomena in dense neutron matter at a strong magnetic field}

\author{A.A. Isayev}
\email{isayev@kipt.kharkov.ua}

\affiliation{%
Kharkov Institute of Physics and Technology, Academicheskaya
Street 1,
 Kharkov, 61108, UKRAINE
\\
Kharkov National University, Svobody Sq., 4, Kharkov, 61077,
UKRAINE }%

\author{J. Yang}
\email{jyang@ewha.ac.kr}

\affiliation{%
Department  of Physics and the Institute for the Early Universe,
 \\
Ewha Womans University, Seoul 120-750, KOREA }

\received{27.10.2009}

\begin{abstract}
Spin polarized states  in neutron matter at  strong magnetic
fields up to $10^{18}$~G are considered in the model with the
Skyrme effective interaction. Analyzing the self-consistent
equations at zero temperature, it is shown that a
thermodynamically stable branch of solutions for the spin
polarization parameter as a function of density corresponds to the
negative spin polarization  when the majority of neutron spins are
oriented oppositely to the direction of the magnetic field.
Besides, it is found that
 in a strong magnetic field the state with the positive spin polarization can be realized as a
metastable state at the high density region in neutron matter. At
finite temperature, the entropy of the thermodynamically stable
branch demonstrates the unusual behavior being larger than that
for the nonpolarized state (at vanishing magnetic field) above
certain critical density which is caused by the dependence of the
entropy on the effective masses of neutrons in a spin polarized
state.
\end{abstract}



\maketitle

\renewcommand{\thefootnote}{\fnsymbol{footnote}}

\renewcommand{\thefootnote}{\roman{footnote}}


\section{Introduction. Basic Equations}
\label{introduction}

Neutron stars observed in nature are magnetized objects with the
magnetic field strength at the surface in the range
$10^{9}$-$10^{13}$~G~\cite{LGS}. For a special class of  neutron
stars such as soft gamma-ray  repeaters  and anomalous X-ray
pulsars, the field strength can be much larger and is estimated to
be  about $10^{14}$-$10^{15}$~G~\cite{TD}. These strongly
magnetized objects are called magnetars~\cite{DT} and comprise
about $10\%$ of the whole population of neutron stars~\cite{K}.
However, in the interior of a magnetar the magnetic field strength
may be even larger, reaching the values about
$10^{18}$~G~\cite{CBP,BPL}. Under such circumstances, the issue of
interest is the behavior of a neutron star matter in a strong
magnetic field~\cite{CBP,BPL,CPL,PG}. Further we will approximate
the neutron star matter by  pure neutron matter as was done, e.g.,
in the recent study~\cite{PG}. As a framework for consideration,
we choose a Fermi liquid approach for description of nuclear
matter~\cite{AIP,IY3} and as a potential of nucleon-nucleon
interaction, we utilize  the Skyrme effective forces.

 The normal (nonsuperfluid) states of neutron matter are described
  by the normal distribution function of neutrons $f_{\kappa_1\kappa_2}=\mbox{Tr}\,\varrho
  a^+_{\kappa_2}a_{\kappa_1}$, where
$\kappa\equiv({\bf{p}},\sigma)$, ${\bf p}$ is momentum, $\sigma$ is
the projection of spin on the third axis, and $\varrho$ is the
density matrix of the system~\cite{I,IY,IY1}. Further it will be
assumed that the third axis is directed along the external magnetic
field $\bf{H}$.  The self-consistent matrix equation for determining
the distribution function $f$ follows from the minimum condition of
the thermodynamic potential~\cite{AIP} and is
  \begin{eqnarray}
 f=\left\{\mbox{exp}(Y_0\varepsilon+
Y_4)+1\right\}^{-1}\equiv
\left\{\mbox{exp}(Y_0\xi)+1\right\}^{-1}.\label{2}\end{eqnarray}
Here the single particle energy $\varepsilon$ and the quantity
$Y_4$ are matrices in the space of $\kappa$ variables, with
$Y_{4\kappa_1\kappa_2}=Y_{4}\delta_{\kappa_1\kappa_2}$, $Y_0=1/T$,
and $ Y_{4}=-\mu_0/T$  being
 the Lagrange multipliers, $\mu_0$ being the chemical
potential of  neutrons, and $T$  the temperature. Given the
possibility for alignment of  neutron spins along or oppositely to
the magnetic field $\bf H$, the normal distribution function of
neutrons and single particle energy can be expanded in the Pauli
matrices $\sigma_i$ in spin
space
\begin{align} f({\bf p})&= f_{0}({\bf
p})\sigma_0+f_{3}({\bf p})\sigma_3,\label{7.2}\\
\varepsilon({\bf p})&= \varepsilon_{0}({\bf
p})\sigma_0+\varepsilon_{3}({\bf p})\sigma_3.
 \nonumber
\end{align}

Using Eqs.~\p{2} and \p{7.2}, one can express evidently the
distribution functions $f_{0},f_{3}$
 in
terms of the quantities $\varepsilon$: \begin{align}
f_{0}&=\frac{1}{2}\{n(\omega_{+})+n(\omega_{-}) \},\label{2.4}
 \\
f_{3}&=\frac{1}{2}\{n(\omega_{+})-n(\omega_{-})\}.\nonumber
 \end{align} Here $n(\omega)=\{\exp(Y_0\omega)+1\}^{-1}$ and
 \begin{align}
\omega_{\pm}&=\xi_{0}\pm\xi_{3},\label{omega}\\
\xi_{0}&=\varepsilon_{0}-\mu_{0},\;
\xi_{3}=\varepsilon_{3}.\nonumber\end{align}

As follows from the structure of the distribution functions $f$,
the quantities $\omega_{\pm}$ play the role of the quasiparticle
spectrum and  correspond to neutrons with spin up and spin down.
The distribution functions $f$ should satisfy the norma\-lization
conditions
\begin{align} \frac{2}{\cal
V}\sum_{\bf p}f_{0}({\bf p})&=\varrho,\label{3.1}\\
\frac{2}{\cal V}\sum_{\bf p}f_{3}({\bf
p})&=\varrho_\uparrow-\varrho_\downarrow\equiv\Delta\varrho.\label{3.2}
 \end{align}
 Here $\varrho=\varrho_{\uparrow}+\varrho_{\downarrow}$ is the total density of
 neutron matter, $\varrho_{\uparrow}$ and $\varrho_{\downarrow}$  are the neutron number densities
 with spin up and spin down,
 respectively. The
quantity $\Delta\varrho$  may be regarded as the neutron spin
order parameter. It determines the magnetization of the system
$M=\mu_n \Delta\varrho$, $\mu_n$ being the neutron magnetic
moment. The magnetization may contribute to the internal magnetic
field $B=H+4\pi M$. However, we will assume, analogously to
Refs.~\cite{PG,BPL}, that the contribution of the magnetization
 to the magnetic field
$B$ remains small for all relevant densities and magnetic field
strengths, and, hence, $B\approx H$. This assumption holds true
due to the tiny value of the neutron magnetic moment
$\mu_n=-1.9130427(5)\mu_N\approx-6.031\cdot10^{-18}$
MeV/G~\cite{A} ($\mu_N$ being the nuclear magneton)
 and is confirmed numerically in a subsequent integration of the self-consistent equations.

In order to get the self--consistent equations for the components
of the single particle energy, one has to set the energy
functional of the system. In view of the above approximation, it
reads~\cite{IY}
\begin{align} E(f)&=E_0(f,H)+E_{int}(f)+E_{field},\label{enfunc} \\
{E}_0(f,H)&=2\sum\limits_{ \bf p}^{} \varepsilon_0({\bf
p})f_{0}({\bf p})-2\mu_n H\sum\limits_{ \bf p}^{} f_{3}({\bf
p}),\nonumber
\\ {E}_{int}(f)&=\sum\limits_{ \bf p}^{}\{
\tilde\varepsilon_{0}({\bf p})f_{0}({\bf p})+
\tilde\varepsilon_{3}({\bf p})f_{3}({\bf p})\},\nonumber\\
E_{field}&=\frac{H^2}{8\pi}\cal V,\nonumber\end{align} where
\begin{align}\tilde\varepsilon_{0}({\bf p})&=\frac{1}{2\cal
V}\sum_{\bf q}U_0^n({\bf k})f_{0}({\bf
q}),\;{\bf k}=\frac{{\bf p}-{\bf q}}{2}, \label{flenergies}\\
\tilde\varepsilon_{3}({\bf p})&=\frac{1}{2\cal V}\sum_{\bf
q}U_1^n({\bf k})f_{3}({\bf q}).\nonumber
\end{align}
Here  $\varepsilon_0({\bf p})=\frac{{\bf p}^{\,2}}{2m_{0}}$ is the
free single particle spectrum, $m_0$ is the bare mass of a
neutron, $U_0^n({\bf k}), U_1^n({\bf k})$ are the normal Fermi
liquid (FL) amplitudes, and
$\tilde\varepsilon_{0},\tilde\varepsilon_{3}$ are the FL
corrections to the free single particle spectrum. Note that in
this study we will not be interested in the total energy density
and pressure in the interior of a  neutron star. By this reason,
the field contribution $E_{field}$, being the energy of the
magnetic field in the absence of matter, can be omitted. Using
Eq.~\p{enfunc}, one can get the self-consistent equations in the
form~\cite{IY} \begin{align}\xi_{0}({\bf p})&=\varepsilon_{0}({\bf
p})+\tilde\varepsilon_{0}({\bf p})-\mu_0, \label{14.2}\\
\xi_{3}({\bf p})&=-\mu_nH+\tilde\varepsilon_{3}({\bf p}).\nonumber
\end{align}

   To obtain
 numerical results, we  utilize the  effective Skyrme interaction.
 The normal FL
amplitudes can be expressed in terms of the Skyrme
  force parameters~\cite{AIP,IY3}:
\begin{align} U_0^n({\bf k})&=2t_0(1-x_0)+\frac{t_3}{3}\varrho^\beta(1-x_3)\label{101}\\ &\quad 
+\frac{2}{\hbar^2}[t_1(1-x_1)+3t_2(1+x_2)]{\bf k}^{2},
\nonumber\\
U_1^n({\bf
k})&=-2t_0(1-x_0)-\frac{t_3}{3}\varrho^\beta(1-x_3)+\frac{2}{\hbar^2}\cdot\label{102}\\ & 
\cdot[t_2(1+x_2)-t_1(1-x_1)]{\bf k}^{2}\equiv a_n+b_n{\bf
k}^{2}.\nonumber\end{align} Further we do not take into account
the effective tensor forces, which lead to coupling of the
momentum and spin degrees of freedom, and, correspondingly, to
anisotropy in the momentum dependence of FL amplitudes with
respect to the spin quantization axis. Then
\begin{align}
\xi_{0}&=\frac{p^2}{2m_{n}}-\mu,\label{4.32}\\
\xi_{3}&=-\mu_nH+(a_n+b_n\frac{{\bf
p}^{2}}{4})\frac{\Delta\varrho}{4}+\frac{b_n}{16}\langle {\bf
q}^{2}\rangle_{3}, \label{4.33}
\end{align}
where  the effective neutron mass  $m_{n}$ reads \begin{align}
\frac{\hbar^2}{2m_{n}}=\frac{\hbar^2}{2m_0}+\frac{\varrho}{8}
[t_1(1-x_1)+3t_2(1+x_2)],\label{181}\end{align} and the
renormalized chemical potential $\mu$ should be determined from
Eq.~\p{3.1}. The quantity $\langle {\bf q}^{2}\rangle_{3}$ in
Eq.~\p{4.33}  is the second order moment of the distribution
function $f_3$:
\begin{align} \langle {\bf
q}^{2}\rangle_{3}&=\frac{2}{V}\sum_{\bf q}{\bf q}^2f_{3}({\bf
q}).\label{6.11}\end{align}  In view of Eqs.~\p{4.32}, \p{4.33},
 the branches $\omega_\pm\equiv\omega_\sigma$ of the quasiparticle spectrum
in Eq.~\p{omega} read \begin{equation}
\omega_\sigma=\frac{p^2}{2m_{\sigma}}-\mu+\sigma\bigl(-\mu_nH+\frac{a_n\Delta\varrho}{4}
+\frac{b_n}{16}\langle {\bf
q}^{2}\rangle_{3}\bigr),\label{spectrud}\end{equation} where
$m_\sigma$ is the effective  mass of a neutron with spin up
($\sigma=+1$) and spin down ($\sigma=-1$) \begin{align}
\frac{\hbar^2}{2m_{\sigma}}&=\frac{\hbar^2}{2m_0}
+\frac{\varrho_\sigma}{2}
t_2(1+x_2)+\frac{\varrho_{-\sigma}}{4}\cdot\label{187}\\&\quad\cdot[t_1(1-x_1)+t_2(1+x_2)],\;
\varrho_{+(-)}\equiv\varrho_{\uparrow(\downarrow)}.\nonumber\end{align}

 Thus, with account of expressions
\p{2.4}  for the distribution functions $f$, we obtain the
self--consistent equations \p{3.1}, \p{3.2}, and \p{6.11} for the
effective chemical potential $\mu$,
  spin  order parameter
$\Delta\varrho$,
 and  second order moment
$\langle {\bf q}^{2}\rangle_{3}$. To check the thermodynamic
stability of different solutions of the self-consistent equations,
it is necessary to compare the corresponding free energies
$F=E-TS$, where  the entropy reads \begin{align} S&=-\sum_{\bf
p}\sum_{\sigma=\uparrow,\,\downarrow}\{n(\omega_{\sigma})\ln
n(\omega_{\sigma})\label{entr}\\ &\quad+\bar n(\omega_{\sigma})\ln
\bar n(\omega_{\sigma})\}, \;\bar n(\omega)=1-n(\omega).\nonumber
\end{align}

\section{Analysis  of the self-consistent equations}

 In solving numerically  the
self-consistent equations, we utilize SLy7 Skyrme
force~\cite{CBH}, constrained originally to reproduce the results
of microscopic neutron matter calculations. We consider magnetic
fields up to the values allowed by the scalar virial theorem. For
a neutron star with the mass $M$ and radius $R$, equating the
magnetic field energy $E_H\sim (4\pi R^3/3)(H^2/8\pi)$ with the
gravitational binding energy $E_G\sim GM^2/R$, one gets the
estimate $H_{max}\sim\frac{M}{R^2}(6G)^{1/2}$. For a typical
neutron star with $M=1.5M_{\odot}$ and $R=10^{-5}R_\odot$, this
yields for the maximum value of the magnetic field strength
$H_{max}\sim10^{18}$~G. This magnitude can be expected in the
interior of a magnetar while recent observations report the
surface values up to $H\sim 10^{15}$~G~\cite{IShS}.

\begin{figure}[tb]
\begin{center}
\includegraphics[bb= 0 0 257 208,width=0.98\linewidth]{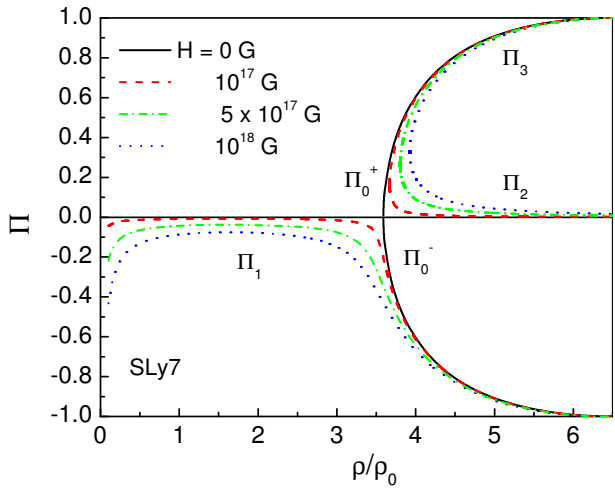} 
\end{center}
\vspace{-2ex} \caption{(Color online) Neutron spin polarization
parameter as a function of density at $T=0$ and different magnetic
field strengths for SLy7 interaction. The branches of spontaneous
polarization $\Pi_0^-,\Pi_0^+$ are shown by solid curves.}
\label{fig1}\vspace{-0ex}
\end{figure}

Fig.~\ref{fig1} shows the neutron spin polarization parameter
$\Pi=\Delta\varrho/\varrho$  as a function of density for a set of
fixed values of the magnetic field at zero temperature. At $H=0$,
the self-consistent equations
 are
invariant with respect to the global flip of neutron spins and we
have two branches  of solutions for the spin polarization
parameter, $\Pi_0^+(\varrho)$ (upper) and $\Pi_0^-(\varrho)$
(lower) which differ only by sign,
$\Pi_0^+(\varrho)=-\Pi_0^-(\varrho)$. At $H\not=0$, the
self-consistent equations  lose the invariance with respect to the
global flip of the spins and, as a consequence, the  branches of
spontaneous polarization are  modified differently by the magnetic
field. The lower branch $\Pi_1(\varrho)$, corresponding to the
negative spin polarization, extends down to the very low
densities. There are three characteristic density domains for this
branch. At  low densities $\varrho\lesssim 0.5\varrho_0$, the
absolute value of the spin polarization parameter increases with
decreasing density. At intermediate densities
$0.5\varrho_0\lesssim\varrho\lesssim3\varrho_0$, there is a
plateau in the $\Pi_1(\varrho)$  dependence, whose characteristic
value depends on $H$, e.g., $\Pi_1\approx-0.08$ at $H=10^{18}$~G.
At densities $\varrho\gtrsim3\varrho_0$,  the magnitude of the
spin polarization parameter increases with density, and neutrons
become totally polarized at $\varrho\approx6\varrho_0$.

 It is seen also from Fig.~\ref{fig1} that beginning
from some threshold density the self-consistent equations at a
given density have two positive solutions  for the spin
polarization parameter (apart from one negative solution). These
solutions belong to two branches, $\Pi_2(\varrho)$ and
$\Pi_3(\varrho)$, characterized by the different dependence from
density. For the branch $\Pi_2(\varrho)$, the spin polarization
parameter decreases with density and tends to zero value while for
the  branch $\Pi_3(\varrho)$ it increases with density and is
saturated. These branches appear step-wise at the same threshold
density $\varrho_{\rm th}$ dependent on the magnetic field and
being larger than the critical density of spontaneous spin
instability in neutron matter.  For example, for SLy7 interaction,
$\varrho_{\rm th}\approx 3.80\,\varrho_0$ at $H=5\cdot 10^{17}$~G,
and  $\varrho_{\rm th}\approx 3.92\,\varrho_0$ at $H= 10^{18}$~G.
The magnetic field, due to the negative value of the neutron
magnetic moment, tends to orient the neutron spins oppositely to
the magnetic field direction. As a result, the spin polarization
parameter for the branches $\Pi_2(\varrho)$, $\Pi_3(\varrho)$ with
the positive spin polarization is smaller than that for the branch
of spontaneous polarization $\Pi_0^+$, and, vice versa, the
magnitude of the spin polarization parameter for the branch
$\Pi_1(\varrho)$ with the negative spin polarization is larger
than the corresponding value for the branch of spontaneous
polarization $\Pi_0^-$. Note that the impact of even such strong
magnetic field as $H=10^{17}$~G is small: The spin polarization
parameter for all three branches $\Pi_1(\varrho)$-$\Pi_3(\varrho)$
is either close to zero, or close to its value in the state with
spontaneous polarization, which  is governed by the spin-dependent
medium correlations.

\begin{figure}[tb]
\begin{center}
\includegraphics[bb= 0 0 256 203,width=\linewidth]{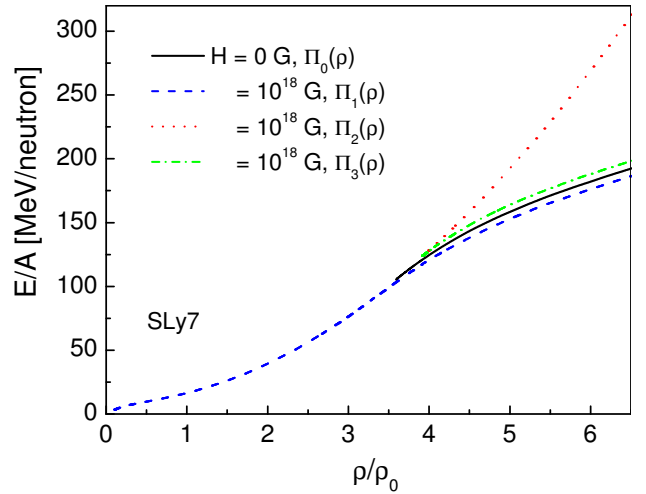} 
\end{center}
\vspace{-2ex} \caption{(Color online) Energy  per neutron  as a
function of density at $T=0$  for different branches
$\Pi_1(\varrho)$-$\Pi_3(\varrho)$ of solutions of the
self-consistent equations at $H=10^{18}$~G, including a
spontaneously polarized state.} \label{fig2}\vspace{-3ex}
\end{figure}

Thus, at densities larger than $\varrho_{\rm th}$, we have three
branches of solutions: one of them, $\Pi_1(\varrho)$,  with the
negative spin polarization and two others, $\Pi_2(\varrho)$ and
$\Pi_3(\varrho)$, with the positive polarization. In order to
clarify, which branch is thermodynamically preferable, one should
compare the corresponding free energies. Fig.~\ref{fig2} shows the
energy per neutron as a function of density at $T=0$ and
$H=10^{18}$~G for these three branches, compared with the energy
per neutron for a spontaneously polarized state [the branches
$\Pi_0^\pm(\varrho)$]. It is seen that the state with the majority
of neutron spins  oriented oppositely to the direction of the
magnetic field [the branch $\Pi_1(\varrho)$] has a lowest energy.
 However, the state, described by the branch
$\Pi_3(\varrho)$ with the positive spin polarization, has the
energy very close to that of the thermodynamically stable state.
This means that despite the presence of a strong magnetic field
$H\sim 10^{18}$~G, the state with the majority of neutron spins
directed  along the magnetic field can be realized as a metastable
state in the dense core of a neutron star in the model
consideration with the Skyrme effective interaction.  Note here
that in the study~\cite{PG} of neutron matter at a strong magnetic
field only thermodynamically stable branch of solutions for the
spin polarization parameter was found in the model with the SLy7
Skyrme interaction.

\begin{figure}[tb]
\begin{center}
\includegraphics[bb= 0 0 259 200,width=\linewidth]{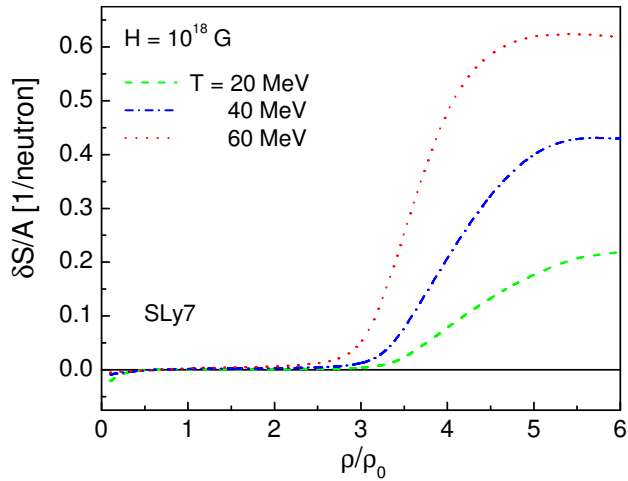} 
\end{center}
\vspace{-2ex} \caption{(Color online) The entropy per neutron for
the $\Pi_1$ branch, measured from its value in the nonpolarized
state (at $H=0$),  as a function of density at $H=10^{18}$ G and
different temperatures. } \label{fig3}\vspace{-0ex}
\end{figure}

One can consider also finite temperature effects on spin polarized
states in neutron matter at a strong magnetic field. Calculations
show that the influence of finite temperatures on spin
polarization remains moderate in the Skyrme model, at least, for
temperatures relevant for protoneutron stars (up to 60~MeV). An
unexpected moment appears when we consider the behavior of the
entropy of spin polarized state as a function of density.
Fig.~\ref{fig3} shows the density dependence of the difference
between the entropies per neutron of the polarized (the $\Pi_1$
branch) and nonpolarized (at $H=0$) states  at different fixed
temperatures. It is seen that with increasing density the
difference of the entropies becomes positive.  It looks like the
polarized state in a strong magnetic field  beginning from some
critical density $\varrho_s$ is less ordered than the nonpolarized
state. Such unusual behavior of the entropy  was found also in the
earlier works for spontaneously polarized states in
neutron~\cite{RPV} and nuclear~\cite{I2,I3} matter with the Skyrme
and Gogny effective forces, respectively. Providing the low
temperature expansion for the entropy in Eq.~\p{entr}, one can get
the condition for the difference between the entropies per neutron
of the polarized and nonpolarized states to be negative in the
form

\begin{equation} \frac{m_{\uparrow}}{m_n}(1+\Pi)^\frac{1}{3}+
\frac{m_{\downarrow}}{m_n}(1-\Pi)^\frac{1}{3}-2<0.\label{9}\end{equation}
\vspace{-1mm}

For low temperatures, it can be checked numerically that this
condition is violated for the $\Pi_1$ branch of the spin
polarization parameter above the critical density $\varrho_s$
being weakly dependent on temperature.


J.Y. was supported  by  grant R32-2008-000-10130-0 from WCU
project of MEST and NRF through Ewha Womans University.

\label{last}

\begin{thebibliography}{99}
\bibitem{LGS} A. Lyne, and F. Graham-Smith, {\it Pulsar Astronomy} (Cambridge
Univ. Press, Cambridge, 2005).
\bibitem{TD} C. Thompson, and R.C. Duncan,   Astrophys. J. {\bf 473},
322 (1996).
\bibitem{DT} R.C. Duncan, and C. Thompson,
 Astrophys. J. {\bf
392}, L9 (1992).
\bibitem{K} C. Kouveliotou,  et al.,   Nature, {\bf 393}, 235
(1998).
\bibitem{CBP} S. Chakrabarty, D. Bandyopadhyay, and S. Pal, Phys. Rev.
Lett. {\bf 78}, 2898 (1997).
\bibitem{BPL} A. Broderick, M. Prakash, and J. M. Lattimer, Astrophys. J. {\bf 537},
351 (2000).
\bibitem{CPL} C. Cardall, M. Prakash, and J.
M. Lattimer, Astrophys. J. {\bf 554}, 322 (2001).
\bibitem{PG} M. A. Perez-Garcia,  Phys. Rev. C {\bf 77}, 065806 (2008).
\bibitem{AIP} A. I. Akhiezer, A. A. Isayev, S. V. Peletminsky, A. P. Rekalo, and
A. A. Yatsenko, JETP {\bf 85}, 1 (1997).
\bibitem{IY3}  A.A.  Isayev, and  J. Yang, in {\it Progress in Ferromagnetism
Research}, edited by V.N. Murray (Nova Science Publishers, New York,
2006), p. 325 [arXiv:nucl-th/0403059].
\bibitem{I} A.A.  Isayev,    JETP Letters {\bf 77}, 251 (2003).
\bibitem{IY}  A.A.  Isayev, and  J. Yang,   Phys.  Rev.  C {\bf 69}, 025801
(2004).
\bibitem{IY1} A.A. Isayev,  Phys.  Rev.  C {\bf 74}, 057301 (2006). 
\bibitem{A}
C. Amsler et al. (Particle Data Group), Phys. Lett. {\bf B667}, 1
(2008).
\bibitem{CBH} E. Chabanat, P. Bonche, P. Haensel, J. Meyer, and R.
Schaeffer, Nucl. Phys. {\bf A635}, 231 (1998).
\bibitem{IShS}A. I. Ibrahim, S. Safi-Harb, J. H. Swank, et al., 
Astrophys. J. {\bf 574}, L51 (2002).
\bibitem{RPV} A. Rios,  A. Polls, and I. Vida$\tilde{\mbox n}$a,   Phys.  Rev.  C {\bf 71},
 055802 (2005).
\bibitem{I2} A.A.  Isayev, Phys.  Rev.  C {\bf 72}, 014313
 (2005).
\bibitem{I3} A.A.  Isayev, Phys.  Rev.  C  {\bf 76}, 047305 (2007).

\end{thebibliography}
\end{document}